 \documentclass[preprint2]{aastex}

\usepackage{epsfig}
\usepackage{multirow}

\shortauthors{Pavlov, Chang, Kargaltsev} 
\shorttitle{Extended X-ray emission from the PSR B1259$-$63/SS\,2883 binary}

\begin{document}

\title{Extended Emission from the PSR B1259-63/SS 2883 Binary Detected
with Chandra}

\author{George G.\ Pavlov\altaffilmark{1,2}, Chulhoon Chang\altaffilmark{1}, and
Oleg Kargaltsev\altaffilmark{3}}

\altaffiltext{1}{Department of Astronomy \& Astrophysics, Pennsylvania State University, PA 16802, USA; pavlov@astro.psu.edu, chchang@astro.psu.edu }
\altaffiltext{2}{St.-Petersburg State Polytechnical University, Polytekhnicheskaya ul., 29, 195251, Russia}
\altaffiltext{2}{Department of Astronomy, University of Florida, FL 32611, USA;
kargaltsev@astro.ufl.edu}

\begin{abstract}
PSR B1259$-$63 is a middle-aged radio pulsar ($P=48$ ms, $\tau=330$ kyr,
$\dot{E}=8.3\times 10^{35}$ erg s$^{-1}$) 
in an eccentric binary ($P_{\rm orb}=3.4$ yr, $e=0.87$) with a high-mass Be 
companion, SS 2883. We observed the binary near apastron with the 
{\sl Chandra} ACIS detector on 2009 May 14 for 28 ks.
In addition to the previously studied pointlike 
source at the pulsar's position, 
 we detected
 extended emission 
on the south-southwest side of this source.
The 
pointlike source spectrum can be 
 described by the absorbed power-law model
with the hydrogen column density 
$N_{\rm H}=(2.5\pm 0.6)\times10^{21}\, {\rm cm}^{-2}$, 
photon index $\Gamma = 
1.6\pm 0.1$, 
and luminosity 
$L_{\rm 0.5-8\, keV} 
\approx 1.3\times 10^{33}\,d_3^2$ 
ergs s$^{-1}$,
where $d_3$ is the distance scaled to 3 kpc. 
This emission likely includes 
an unresolved part of
the pulsar wind nebula (PWN) created by the
colliding winds from the pulsar and the Be companion,
and a contribution from the pulsar magnetosphere.
The extended emission
apparently consists of two components.
The highly significant compact component looks like a southward extension
of the pointlike source image,
seen up to $\sim 4''$ from the pulsar position. 
Its spectrum has
about the same slope as the 
pointlike source spectrum, while its 
luminosity is a factor of 10 lower.
We also detected
an elongated feature extended
$\sim 15''$ southwest of the pulsar, 
but significance of this detection is marginal. 
We tentatively interpret the resolved compact PWN component as a
shocked pulsar wind blown out of the binary by the wind of the Be
component, while the elongated component could be a pulsar jet.
\end{abstract}


\section{Introduction}
The outstanding resolution of {\sl Chandra} has made possible detailed
 investigations of pulsar wind nebulae (PWNe). 
PWNe are formed by the relativistic pulsar 
wind (PW), which shocks in the ambient medium and emits synchrotron radiation. 
The PWN morphology depends on the Mach number, 
$\mathcal{M}= v/c_s$, where $v$ is the pulsar's velocity relative to the 
ambient medium, and $c_s$ is the speed of sound. PWNe created by 
subsonically moving pulsars, such as the famous Crab, exhibit ``torus-jet'' 
morphology, where the torus represents a shocked equatorial wind beyond 
the ring-like termination shock (TS), while the jets along the pulsar's 
spin axis are polar outflows. At $\mathcal{M}\gg1$, when the ram pressure 
exceeds the ambient pressure, 
$p_{\rm ram} = \rho_{\rm amb}v^2 \gg p_{\rm abm}$, the TS acquires a 
bullet-like shape \citep{Gaensler2004,Bucciantini2005}, and the PWN exhibits 
a cometary bow-shock morphology, with a long tail behind the moving 
pulsar. Spectacular examples of such PWNe are created by the pulsars 
J1747$-$2958 
\citep{Gaensler2004} and J1509$-$5850 
\citep{Karga2008b}. The tails may coexist with pulsar jets, often
aligned 
with the proper motion direction \citep[e.g.,][]{Pavlov2003}.

An interesting modification of bow-shock PWNe is expected when the 
pulsar is in a 
high-mass binary system. In such a system, the pulsar 
moves in the wind emanating from the high-mass companion, and 
the bow-shock PWN is produced by the collision of the relativistic PW with 
the dense nonrelativistic wind of the companion
(instead of the interstellar medium (ISM) in the case of
 a solitary pulsar). The overall appearance, luminosity, and spectrum 
of such a PWN depend on 
the ratio of the momentum 
fluxes carried by the winds, 
the velocities $\vec{v}_w$ and $\vec{v}_p$ 
of the companion's wind and the pulsar's orbital motion, the temperature and 
density of the companion's wind, the anisotropy of the outflows, and the 
separation between the pulsar and the companion \citep[hereafter TA97]
{Tavani1997}. 
As these parameters vary along the orbit, the PWN's luminosity, spectrum, 
shape, size, and orientation should also vary. If such a PWN can be resolved, 
observations of the variable PWN morphology would allow a  more 
definitive study of the wind properties than observations of an unresolved
PWN + binary system. 

A promising target for such studies is the well-known
 binary system PSR B1259$-$63/SS\,2883.
PSR B1259$-$63 (hereafter B1259) is a middle-aged radio
pulsar, with the period $P=47.8$ ms, spin-down age $\tau=P/2\dot{P}=330$ kyr,
spin-down power $\dot{E}=8.3\times10^{35}$ ergs s$^{-1}$, and magnetic
field $B=3.3\times 10^{11}$ G \citep{Wang2004}. It is in 
an eccentric, wide binary \citep[$P_{\rm bin}=1237\, {\rm d}\approx
3.4\,\,{\rm yr}$, $e=0.87$, $a\,\sin i=3.9\times10^{13}$ cm, 
$i\approx 36{\degr}$;][]{Johnston1992,Johnston1994}
with a 10th-magnitude Be star SS\,2883 
\citep[$M_{\ast}\approx 10M_{\odot}$, $R_{\ast}\approx 6R_{\odot}$;][]{Johnston1994}. 
The distance  estimates based on the dispersion measure (DM = 147 pc cm$^{-3}$)
give $d=4.6$ kpc for 
the model of the Galactic electron density distribution by \citet{Taylor1993},
and $d=2.75^{+0.61}_{-0.49}$ kpc for the NE2001 model by \citet{Cordes2002}.
However, \citet{Johnston1994} place the system at $d\approx 1.5$ kpc,
in the Saggitarius arm, based on the properties of the optical companion.
As the distance remains somewhat uncertain, we will scale it to 3 kpc below,
$d= 3\,d_3$ kpc.

Discovered with {\sl ROSAT} \citep{Cominsky1994}, X-ray emission from B1259
 has been studied with  {\sl ASCA} \citep{Kaspi1995,Hirayama1999}, 
{\sl Beppo-SAX} \citep{Nicastro1998}, {\sl INTEGRAL} \citep{Shaw2004}, 
{\sl XMM-Newton} \citep{Chernyakova2006,Chernyakova2009}, {\sl Suzaku} \citep{Uchiyama2009}, 
and {\sl Swift} and {\sl Chandra} \citep{Chernyakova2009}. TeV emission from this system 
near periastron was detected with the HESS observatory
 \citep{Aharonian2005, Aharonian2009}. 
The multi-year observations have shown that its X-ray flux and spectrum vary 
with orbital phase. For instance, the 1$-$10 keV energy flux is at minimum, 
$1.0\times10^{-12}$ erg cm$^{-2}$ s$^{-1}$ 
(i.e., $L_X=1.1\times10^{33}\,d_{3}^2$ ergs s$^{-1}$)
at apastron, while it is a factor of 20 higher nine days before periastron. 
The slope of its power-law (PL) spectrum also varies, from $\Gamma\approx 1.7$
 (at apastron) to 1.2 (25 days before periastron) to 2.0 (at periastron). 
No pulsations with the radio pulsar period have been detected from the system
outside the radio range. The orbital variations of the X-ray flux and spectrum 
have been explained by a model in which the X-ray emission is generated in a 
PWN formed by the colliding winds; the PWN becomes more compact and luminous 
when the pulsar crosses the equatorial disk outflow (angular half-width 
of about 20$\degr$) of the Be companion, tilted with respect to the orbital 
plane by an angle $\sim$ 25$\degr$ \citep[TA97;][]{Chernyakova2006}. 
There is no general consensus on the X-ray emission  mechanism 
(synchrotron emission or inverse Compton (IC) scattering of companion's 
optical/UV photons). 

It would be interesting to study the unusual PWN of B1259 in X-rays, but,
at realistic wind parameters, the 
characteristic size of the TS surface
($\lesssim$ 0$\farcs$01) in the B1259 PWN
 is too small to resolve it even with {\sl Chandra}.
On the other hand, we know from the models of bow-shock PWNe
and observations of supersonic solitary pulsars that the 
flow speed in
the collimated tail behind the 
moving pulsar can be so high that the length of the observable 
part of the tail
 can exceed the 
characteristic TS size by a very large factor \citep{Karga2008b},
despite the fast synchrotron and IC cooling of the emitting 
relativistic particles. 
The tail 
in the immediate vicinity of the binary should be oriented along the vector 
$\vec{v}_w - \vec{v}_p$, i.e., almost along the line connecting the Be star and 
the pulsar (as $v_w 
\gg v_p$).
The tail should rotate with the binary period, so that its
large-scale appearance would resemble an Archimedes spiral \citep{Dubus2006}.
Optimal binary phases to search for the outflow are near the apastron,
when the pointlike source emission is the faintest, and the size of the bow-shock
PWN is the largest. We should note that the four {\sl Chandra} observations
taken near the periastron of 2007 \citep{Chernyakova2009} were not
suitable for detecting extended emission because they were short (3--7 ks)
and used either Continuous Clocking mode (which produces only 
one-dimensional  image)
or High Energy Transmission Grating.
In addition to the tail outflow, a high-resolution image may show pulsar jets, 
which can penetrate through the companion's wind, and whose orientation 
would not show any orbital dependence.

To search for extended X-ray emission from the PWN tail and pulsar jets, and to better
 understand the nature of stellar and pulsar winds and their interaction, 
we observed B1259 with {\sl Chandra} near apastron of 2009. 
We describe this observation and the
data analysis in Section 2, and discuss our results in Section 3.

\begin{figure}[h!]
\centering
\includegraphics[scale=0.34]{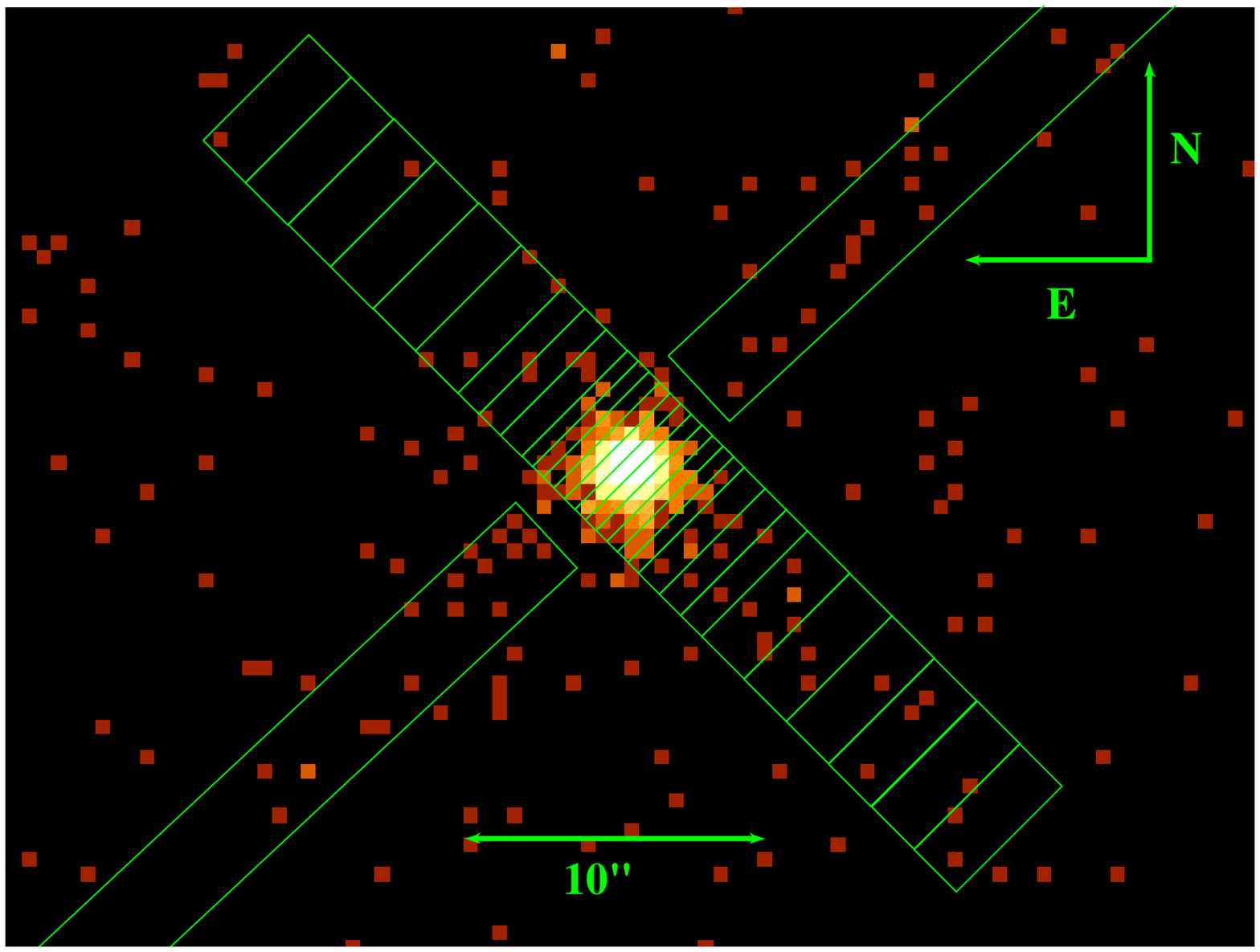}\\
\includegraphics[scale=0.34]{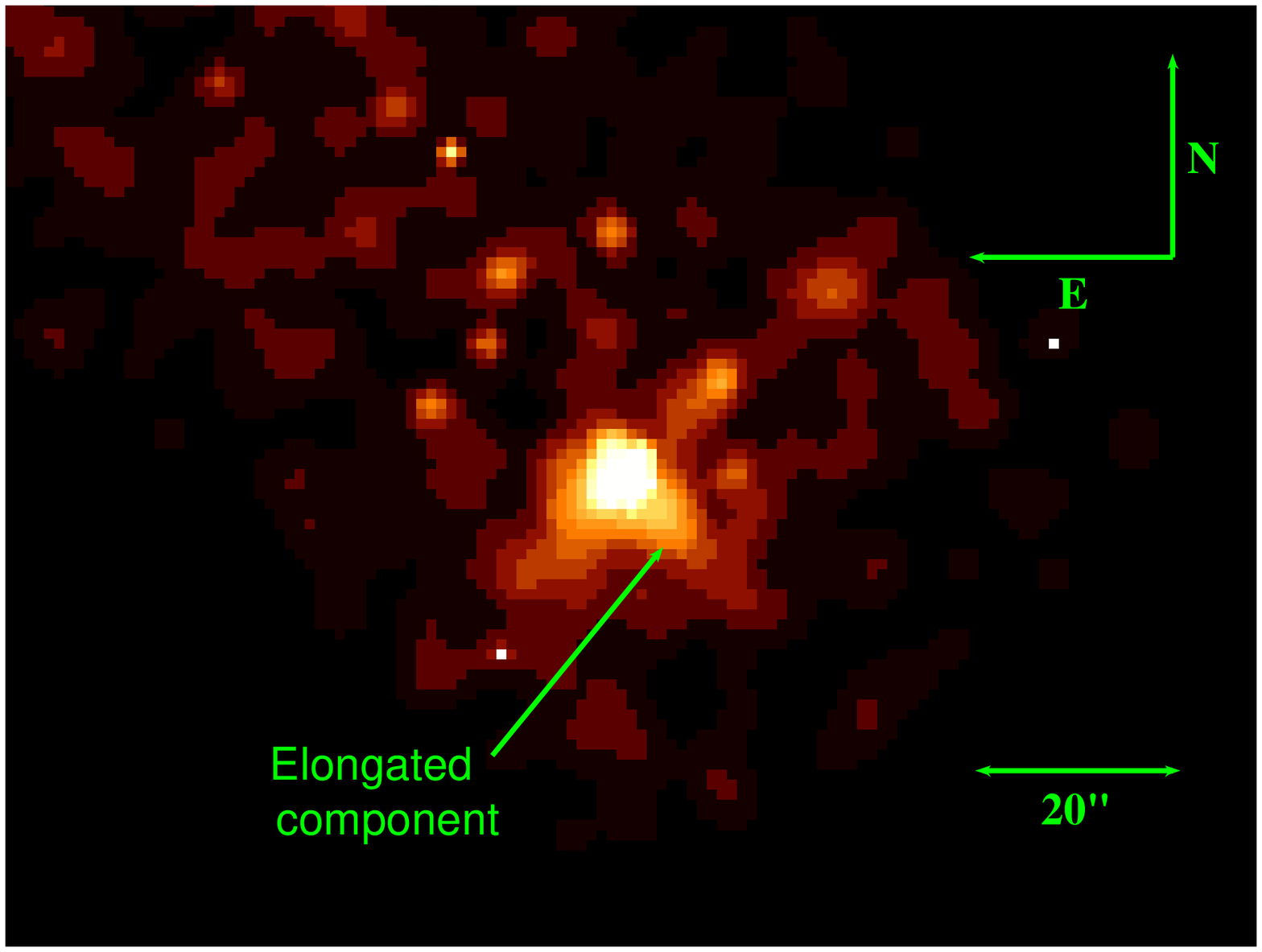}\\
\includegraphics[scale=0.34]{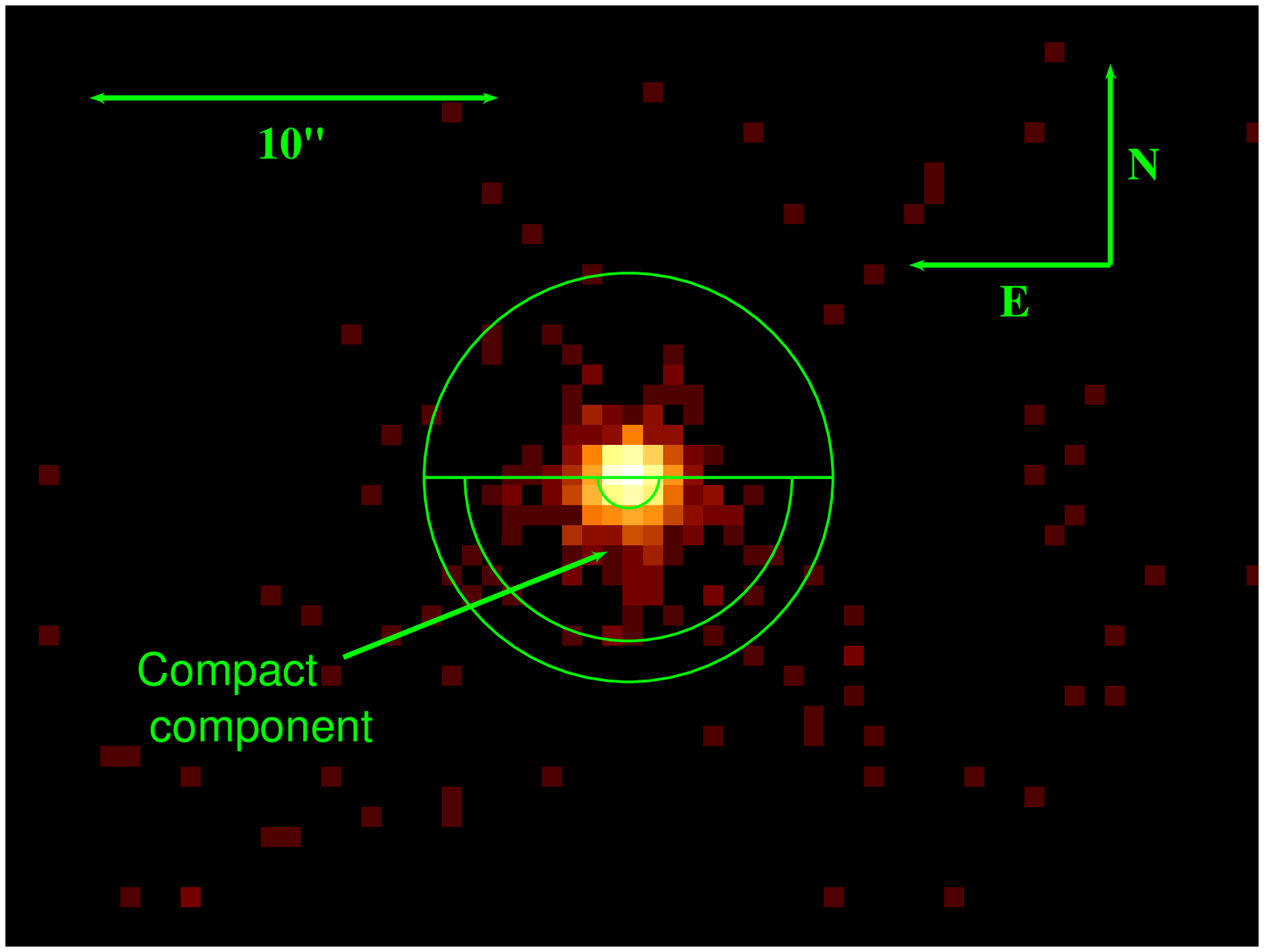}
\caption{
ACIS-I3 images of B1259.
{\it Top}: Original (unbinned, unsmoothed) image
in the 0.5--8 keV band.
The $3''$-wide strips northwest and southeast of the pulsar
are used for the analysis of
the trailed image of the pointlike source.
The $36''\times 5''$ rectangle northeast and southwest of the pulsar
is used for the analysis of extended emission.
{\it Middle:}
 The 0.5--8 keV image
binned to a pixel size
of 0\farcs98 and smoothed with {\it csmooth} to emphasize
the extended emission.
{\it Bottom:} Unbinned, unsmoothed image in the 1--7 keV band;
the semicircles are used for the imaging and spectral
 analyses of extended emission (see text and Table 1).
}
\label{fig1}
\end{figure}

\section{Observations and Data Analysis}
We observed B1259 with the Advanced CCD Imaging Spectrometer (ACIS) 
on board {\sl Chandra} on 2009 May 14 (ObsID 10089),
39 days after apastron (i.e., at the binary phase $\phi=0.532$,
counted from periastron, or the true anomaly $\theta=181.63^\circ$).
The target was observed for 28.28 ks 
in timed exposure mode using ``very faint'' telemetry format. 
To minimize the pile-up effects, we used a custom 124 pixel subarray and turned off all the ACIS chips but I3.
The frame time in this configuration is 0.44104 s (0.4 s exposure time 
and 0.04104 s the image transfer time).
The Y-offset of $-0\farcm2$ and Z-offset of $-0\farcm25$ put the target $29''$ off-axis, at a distance of $>20''$
from the nearest boundary of the $\approx 8'\times 1'$ field of view (FOV),
with allowance for the  $8''$ amplitude dither.
There were no significant flares during the observation. The useful 
scientific exposure time (livetime) was 25.65 ks. The data were reduced and 
analyzed with the {\sl Chandra}\/ Interactive Analysis Observations (CIAO) 
package (ver.\ 4.2), using CALDB 4.2.0. We used {\sl Chandra}\/
Ray Tracer (ChaRT)\footnote{See \url{http://cxc.harvard.edu/chart/threads/index.html}.} and 
MARX\footnote{See \url{http://space.mit.edu/CXC/MARX/}.} software 
for image analysis,  
and XSPEC (ver.\ 12.5.1) for spectral analysis.

\subsection{Image Analysis}
To produce images of the pulsar and its vicinity at subpixel resolution,
we removed the pipeline pixel randomization and applied the
subpixel resolution tool to split-pixel events in the
image \citep{Mori2001,Tsunemi2001}.
Figure~\ref{fig1} shows images of the region around B1259 on the
I3 chip in the 0.5--8 keV and 1--7 keV energy bands.
The coordinates of the bright
compact source at the image center
(2050 counts in the 1\farcs5 radius aperture in the 0.5--8 keV band) are
R.A.(J2000) = $13^{\rm h}02^{\rm m}47\fs625$,
Decl.(J2000) = $-63\arcdeg50\arcmin08\farcs38$,
as measured by the CIAO {\em wavdetect} tool.
This position differs by
0\farcs36 from the radio pulsar position,
R.A.(J2000) = $13^{\rm h}02^{\rm m}47\fs65(1)$,
Decl.(J2000) = $-63\degr50\arcmin08\farcs7(1)$,
for the epoch MJD 50,357 \citep{Wang2004};
such a difference is within the uncertainty of the absolute {\sl Chandra}
astrometry (e.g.,
0\farcs62 at the 90\% confidence level for sources within $2'$
from the optical axis on the ACIS-I3 chip\footnote{See
Figure 5.4 in the {\sl Chandra} Proposer's Observatory Guide (POG),
ver.\ 12 \url{http://cxc.harvard.edu/proposer/POG/}.}).

In the image we see two extended structures apparently associated with the 
pulsar:
a narrow linear structure seen on both sides of the pulsar
along the northwest-southeast direction, and a one-sided south-southwest
extension of pointlike source image.
As the linear structure is oriented along the CCD column 
direction,
it could be naturally interpreted 
as the ``trailed image'' 
of the pointlike source, formed by the source photons arriving during the 
CCD charge transfer to the framestore region. To confirm this explanation
and check that no additional extended emission is required to produce
the observed structure,
we measured the number of counts in a $3''$-wide strip, 
which goes parallel to the chip columns
 from one edge of the subarray to the other 
(excluding a $7''$ gap with the pulsar in the middle;
 see the upper panel of Figure \ref{fig1}),
and compared it with the number of counts expected in the trailed image. 
The background surface brightness, 
0.044 counts arcsec$^{-2}$ in the 0.5--8 keV band, 
was estimated in a $25''$ radius circle
located about $50''$ northeast from the pulsar.
We found 44 counts in the strip,
of which 9.2 counts belong to the
background, i.e., there are $34.8\pm 6.7$ background-subtracted counts.
The number of counts expected in the trailed image can be estimated as
$N_{\rm trail}=C_{\rm src} t_{\rm trail} (n_{\rm strip}/n_{\rm chip})
=22.5\pm 0.6$,
where $C_{\rm src} = 79.9\pm 1.8$
counts ks$^{-1}$ is the
source count rate in the $1\farcs5$ aperture, 
$t_{\rm trail}=2.63$ ks is 
the total image transfer time during the
observation, $n_{\rm strip}= 110$
pixels is the strip length in chip coordinates,
and $n_{\rm chip}=1024$ pixels is the chip size. 
Although the number of counts detected in the strip
 exceeds $N_{\rm trail}$ by 
$12.3\pm 6.7$ counts, the statistical significance of the excess is low
($1.8 \sigma$), which means that the northwest-southeast linear structure
is indeed the trailed image, without a significant contribution from
an extended source.

To evaluate the statistical significance of the other extended feature
(south-southwest of the pulsar, see Figure 1), 
we simulated 
an observation of a point source with the same properties as the observed 
ones using the ChaRT and MARX packages. 
To reduce 
the statistical errors of the simulation, we increased the exposure time 
to 200 ks when 
running ChaRT
and then rescaled the image
to compare with the data. We have also tried several values of the 
DitherBlur parameter of MARX to match the simulated 
point spread function (PSF) with the real data 
and found the best match at the value of $0\farcs2$. 

\begin{figure}[h!]
\centering
\includegraphics[scale=0.3,angle=90]{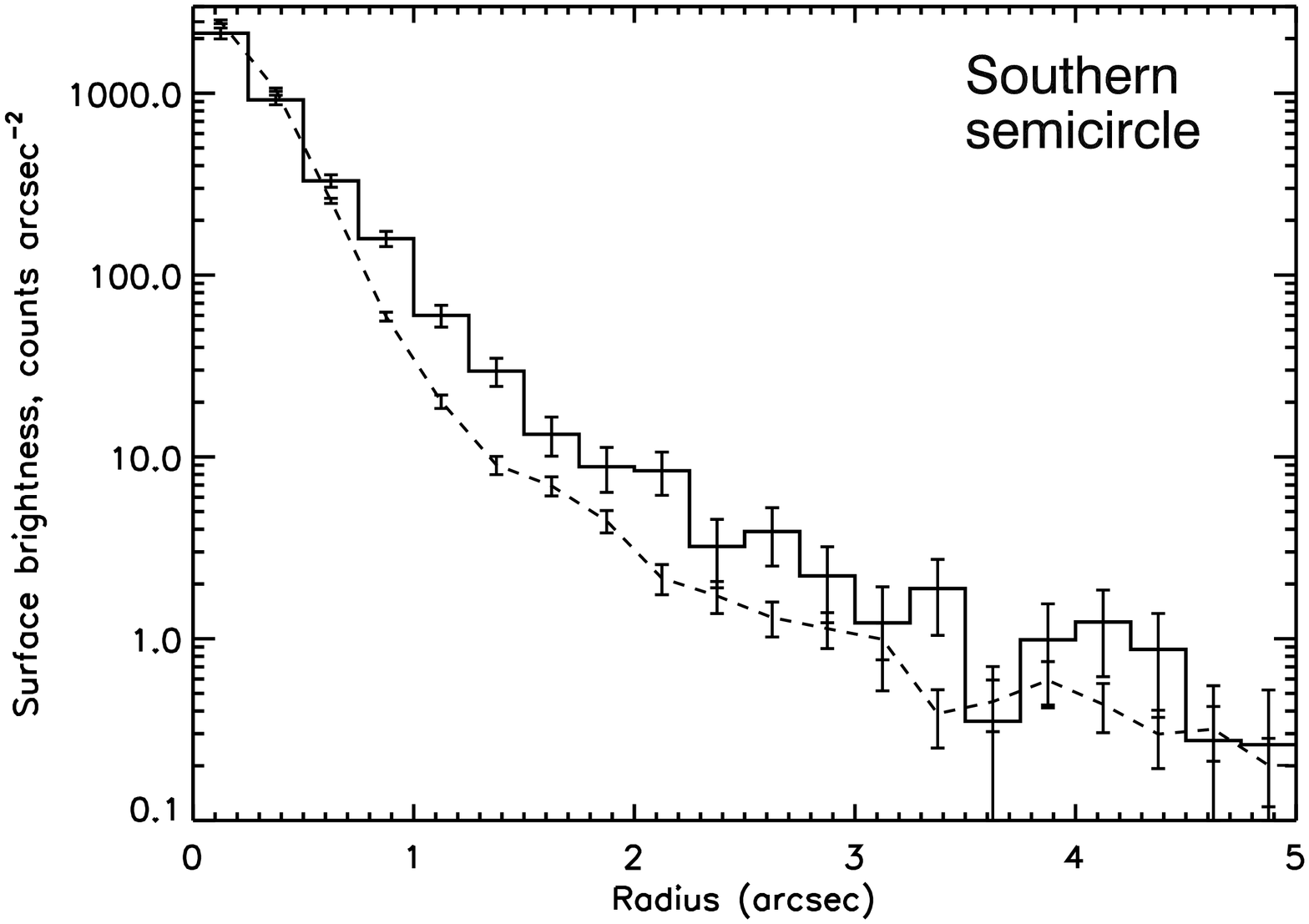}
\includegraphics[scale=0.3,angle=90]{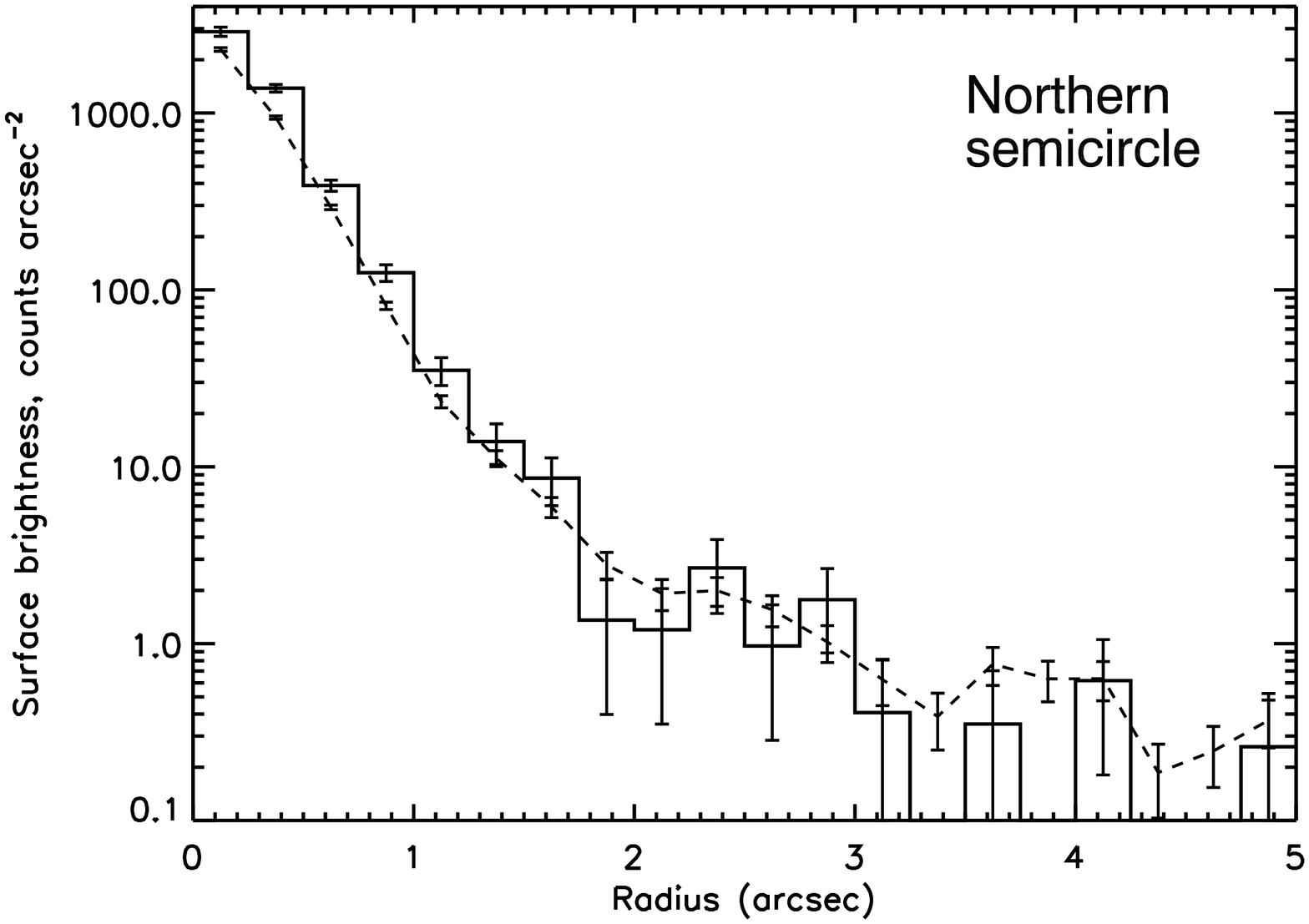}
\caption{
Observed radial count distributions (histograms) in the
southern (top) and northern (bottom) semicircles (see the bottom panel in Figure \ref{fig1}), in the 1--7 keV band, are
 compared with the ChaRT + MARX simulation of a point source
(dashed line).
}
\label{fig2}
\end{figure}

\begin{figure}[ht]
\centering
\includegraphics[scale=0.33,angle=90]{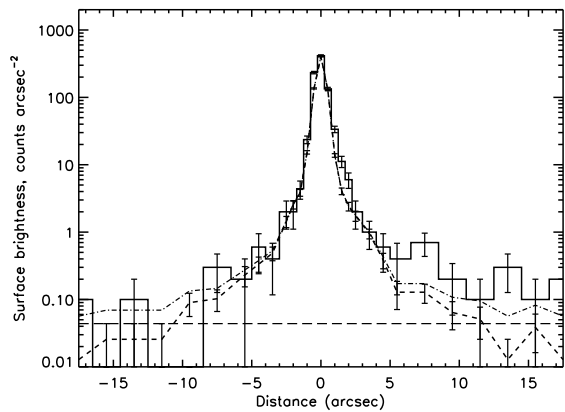}
\caption{Observed
count distribution
(histogram)
along the length of the $36''\times 5''$ rectangle shown in the top panel of
Figure \ref{fig1} (approximately in the northeast-southwest
direction) is compared with the ChaRT+MARX simulation of a point source
(dashed line). The positive and negative  distances correspond to the southwest
and northeast directions.
The horizontal dashed line shows the background level (0.044 counts arcsec$^{-2}$),
 while the dash-dot line corresponds to
 the sum of the background and the simulation.
}
\label{fig3}
\end{figure}

The extended emission apparently consists of a compact part, 
which is seen 
up to about $4''$
 south of the pulsar,
and a narrow elongated structure 
of about $10''$--$15''$
length southwest of the
pulsar, which might bend northward at the apparent end. 
To estimate the significance of these structures, 
we investigated the count distributions in two semicircles of $5''$
radius (north and south of the pulsar; see the bottom panel of Figure~\ref{fig1})
and in the $36''\times 5''$ rectangular region 
(oriented in the northeast-southwest direction, i.e., along the 
elongated structure;
see the top panel of Figure~\ref{fig1}). 

Figure~\ref{fig2} shows the observed and simulated radial distributions 
of counts in the northern and southern semicircles,
in the 1--7 keV band. 
We see that in the northern semicircle
 the observed distribution 
is rather close to the simulated one, while
in the southern semicircle there is a significant excess of the observed counts
with respect to both the simulated distribution and the observed counts
in the northern semicircle.
For instance, for radii between $0\farcs75$ and $4''$, the numbers of counts
in the southern and northern semicircles are 
276 and 160, respectively, i.e.,
the excess is 
$116\pm 21$ counts, significant at the $5.6 \sigma$ level.

Figure~\ref{fig3} shows the distribution of 0.5--8 keV counts along the length $l$
 of the $36''\times 5''$ rectangle 
($l=0$ corresponds to the pulsar position; $l>0$ southwest of the pulsar).
We see an excess of observed counts over the simulated profile
in the southwest half of the rectangle. For instance, at $1''<l<18''$ there are
158 counts;
subtracting the
contribution from the PSF tail and background (65.2 and 3.7 counts, 
respectively), 
we obtain the excess of $89\pm13$ counts,
significant at the $6.9\sigma$ level (if only statistical uncertainties
are taken into account). 
The difference between the numbers of observed counts in 
the rectangle segments symmetric with respect to the pulsar position
confirms the excess. For instance, the difference of
$158 - 91 = 67\pm 16$ counts between the $1''<l<18''$ and $-18'' < l < -1''$
is significant at the $4 \sigma$ level.

The count excess southwest of the pulsar is, however, mostly due to 
the brightening in the immediate pulsar vicinity, at $l\approx 1''$--$2''$, 
while the statistical significance of the
apparent excess at $l>5''$ (the alleged elongated component)
is marginal. For instance,
in the 1--7 keV band, we detected 15 and 3 counts in the 
$5''<l<18''$ and $-18''<l<-5''$ segments, respectively,
so that the difference, $12.0\pm 4.2$ counts, is significant 
at the $2.9 \sigma $ level,
too low to firmly rule out a statistical fluctuation.

The presence of asymmetric extended emission at distances $\gtrsim 1''$
from the pulsar is also supported by Figure 4, which shows the result of
subtraction of the simulated point source image from the observed one.
In addition to the asymmetric extended emission,
the subtracted image  shows some 
azimuthally asymmetric systematic residuals of both signs
within
$\approx 1''$ from the point source position,
resembling those reported recently by 
\citet{JudaKarovska2010} and \citet{Kashyap2010}\footnote{
We should note that the spatial structure of the residuals in Figure 4
differs significantly from that reported by these authors, 
but the structure strongly depends
on the choice of the image center within the ACIS pixel.}. 
This means that the actual PSF core is different from the
 azimuthally symmetric model;
this difference, however, makes no effect beyond $1''$ from the
point source position\footnote{See \url{http://cxc.harvard.edu/ciao/caveats/psf\_artifact.html\#fig1}. }.

\begin{figure}[h!]
\centering
\includegraphics[scale=0.4]{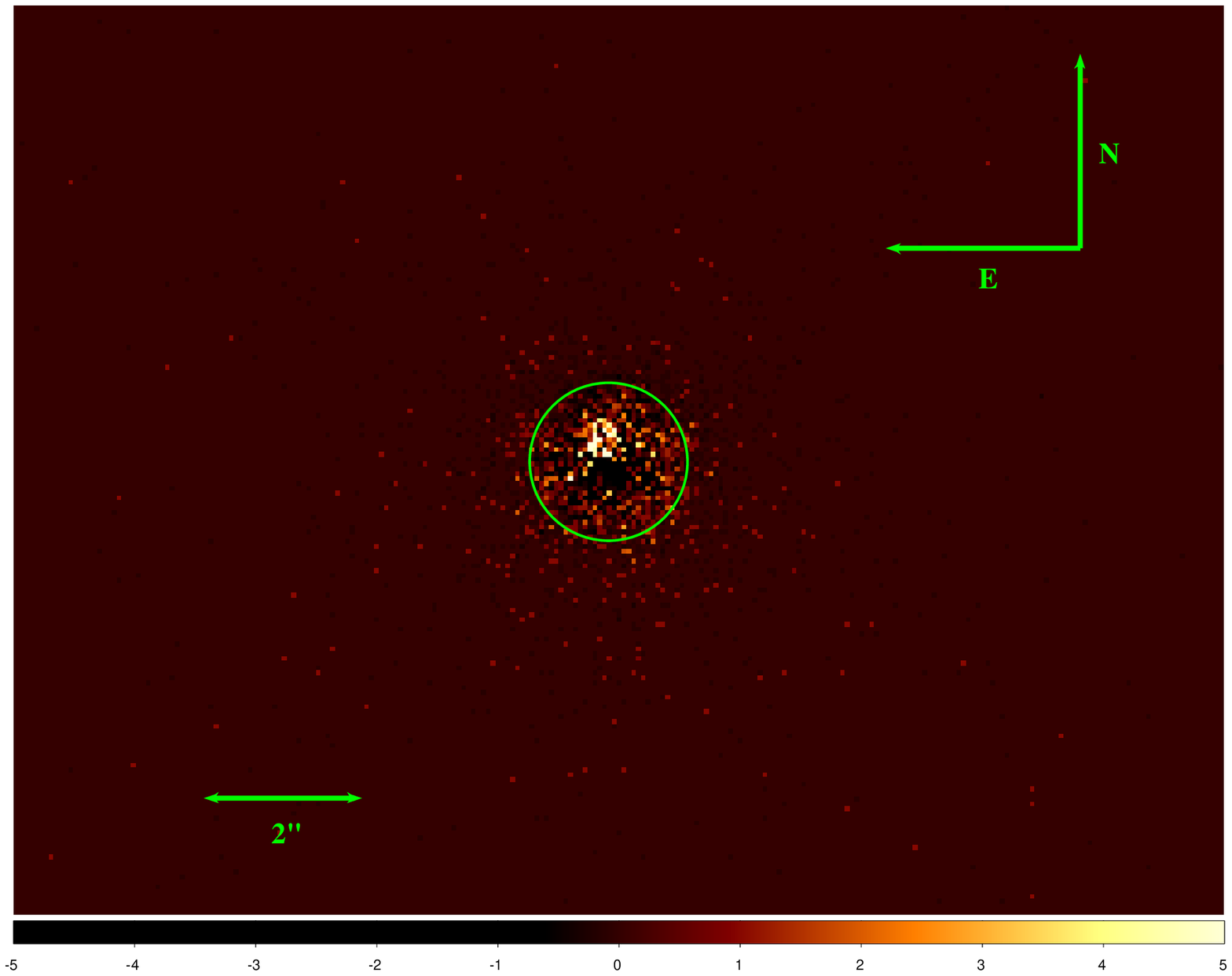}
\includegraphics[scale=0.4]{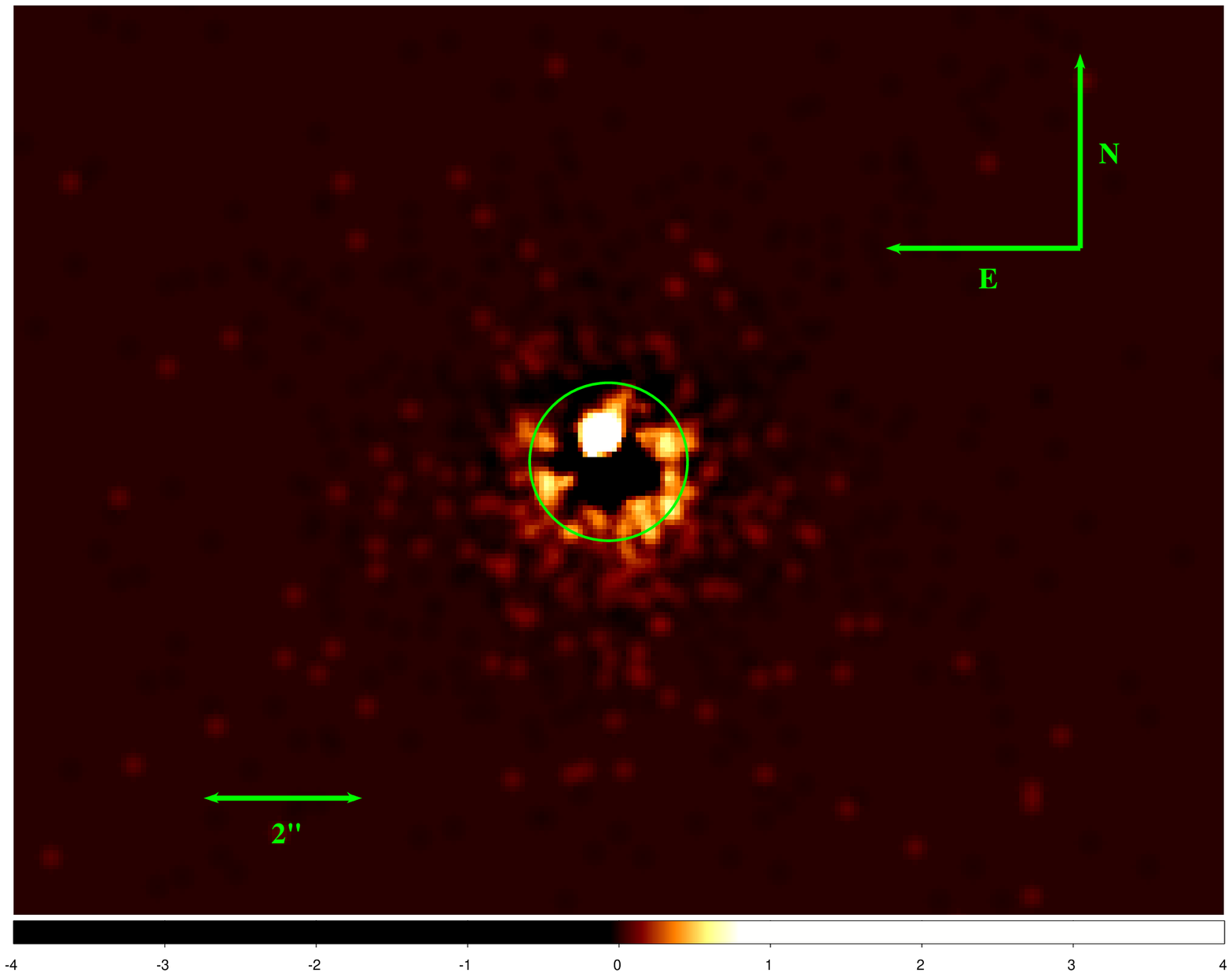}
\caption{{\em Top:} ACIS-I3 image of B1259 with the simulated point source contribution
subtracted. The ChaRT+MARX simulated image (with DitherBlur = 0\farcs2)
 was centered on the {\em wavdetect}
position (see Section 2.1).
 Both the observed and simulated images were rebinned to 1/8 of the
original ACIS pixel size.
The residual structure within the $1''$ radius circle is caused by
imperfections of the simulated PSF core (see Kashyap 2010).
{\em Bottom:} The same image smoothed with the 0\farcs19 Gaussian kernel.
}
\end{figure}

In principle, 
an asymmetry of a point source image can be caused by the tilt 
of the innermost {\sl Chandra} telescope mirrors 
(the mirror pair 6) with respect to the other mirrors \citep[see Section 4.2.3 of POG, and][]{Jerius2004}. It has been observed in bright sources 
\citep[e.g., the Vela pulsar; see][]{Pavlov2001} as an image elongation
between the $-Z$ and $+Y$ directions (in spacecraft coordinates).
In our case, the extended emission is seen between
the $+Z$ and $-Y$ directions. Moreover, the effect of the innermost mirror
misalignment is significant for higher photon energies ($E\gtrsim 5$ keV),
while the asymmetry we observe in the B1259 image is about the same in the
0.5--1.75 and 1.75--8.0 keV images (which have approximately the same numbers of
counts), and our spectrum has too few high-energy counts that could cause
this asymmetry. 
Therefore, the observed one-sided extended emission is not an artefact
caused by the mirror misalignment. 
We suggest that this emission can be interpreted
as an extended PWN of B1259
and conclude that 
the compact component of this PWN, 
up to $\sim 4''$ from the pulsar, is firmly detected, 
while a longer observation is needed to confirm the reality of 
the elongated component.

\begin{figure}[h!]
\centering
\includegraphics[scale=0.34,angle=-90]{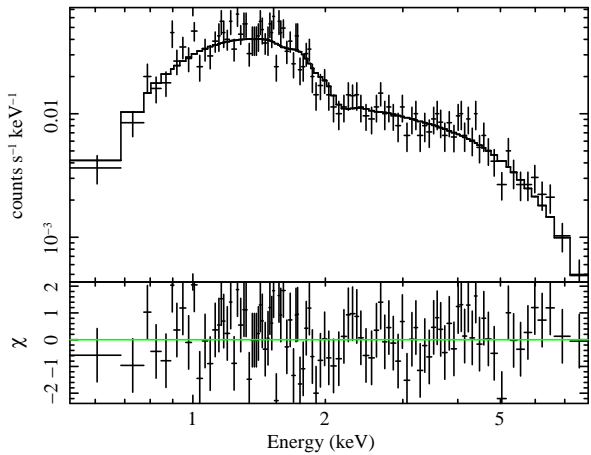}
\caption{Spectrum of the pointlike source
 fitted with the absorbed PL model in the energy range of 0.5--8 keV.
}
\end{figure}

\subsection{Spectral Analysis}
\subsubsection{Pointlike source}
For the spectral analysis of the pointlike source (which likely
includes the unresolved part of the PWN, in addition to the pulsar), 
we extracted 
1874 counts in the 0.5--8 keV band
 from a small circular aperture of $1''$ radius. 
As the encircled count fraction is 88\% for such an aperture and energy band
(according to our ChaRT/MARX simulations), the aperture-corrected 
count rate is 
$83.9\pm 1.9$ counts ks$^{-1}$, or 
0.034 counts per frame,
which corresponds to a 
negligibly small pile-up fraction (about 1\%, according to 
{\sl Chandra} PIMMS\footnote{See \url{http://cxc.harvard.edu/toolkit/pimms.jsp}.}).
The background 
contribution (0.14 counts) is negligible in such a small aperture.
We binned the spectrum with minimum of 15 counts per bin and fit it with 
the absorbed PL model (see Figure 5). 
The fit is good ($\chi^2_{\nu} =0.99$ for 99 degrees of freedom);
its results are reported in Table~\ref{tab1}, while Figure 6 shows the confidence
contours for the fitting parameters.
The fit results are consistent with those obtained  
in an {\sl XMM-Newton} observation
near the 2002 apastron
\citep[$N_{\rm H, 21}=2.9\pm0.2$, $\Gamma=1.69 \pm 0.04$,
$F = (1.04\pm 0.49)\times  10^{-12}$
erg cm$^{-2}$ s$^{-1}$ in the 1--10 keV band\footnote{Because of the relatively low angular
resolution of {\sl XMM-Newton}, those data include both
the pointlike source and the extended emission.}; see][]{Chernyakova2006} as well as with the
{\sl ROSAT} \citep{Cominsky1994} and {\sl ASCA} \citep{Hirayama1999}
near-apastron observations in 1992 and 1995, respectively.

\begin{figure}[h!]
\centering
\includegraphics[scale=0.3,angle=90]{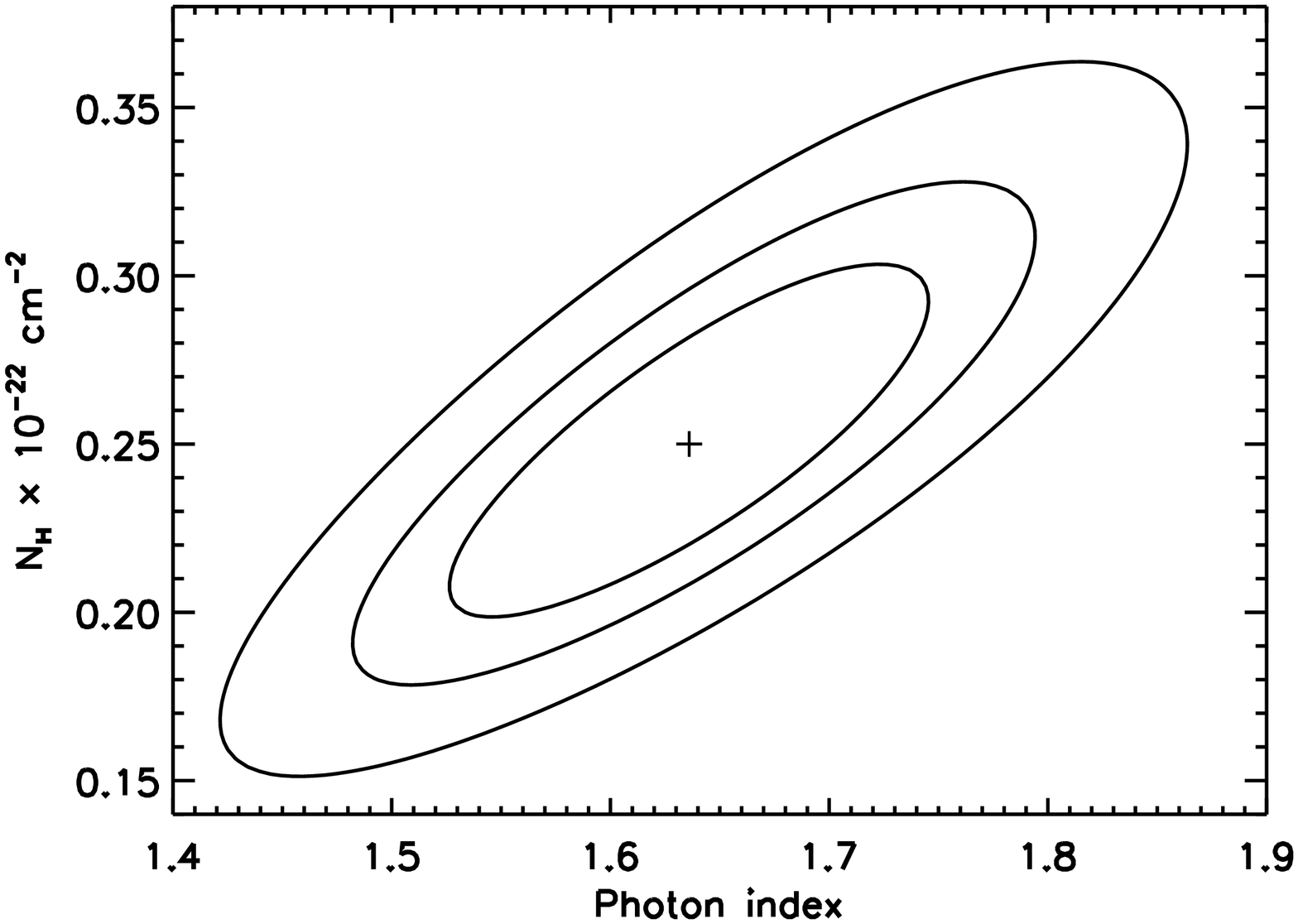}
\includegraphics[scale=0.3,angle=90]{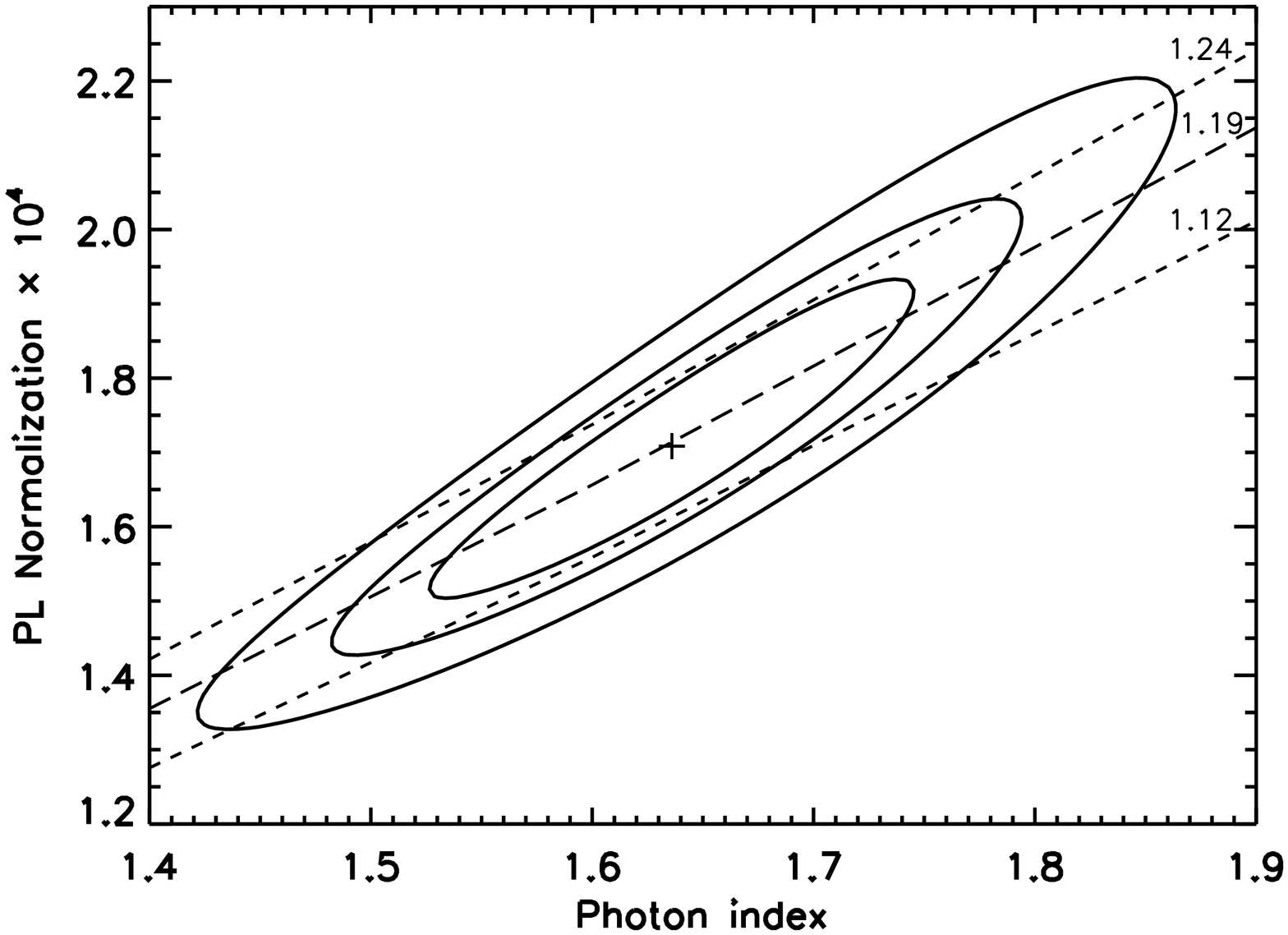}
\caption{
Confidence contours (68\%, 90\%, 99\%) in the
$\Gamma$-$N_{\rm H}$ ({\em top}) and $\Gamma$-${\cal N}$ ({\em bottom}) planes
 for the ACIS spectrum of the pointlike source,
computed for the absorbed PL model.
The PL normalization ${\cal N}$ is in units of
10$^{-4}$ photons cm$^{-2}$ s$^{-1}$ keV$^{-1}$ at 1 keV.
The dashed lines are the lines of constant unabsorbed flux in the 0.5--8 keV
band (the numbers near the lines are the flux values in
 units of $10^{-12}$ erg cm$^{-2}$ s$^{-1}$).
 }
\end{figure}

\begin{figure}[h!]
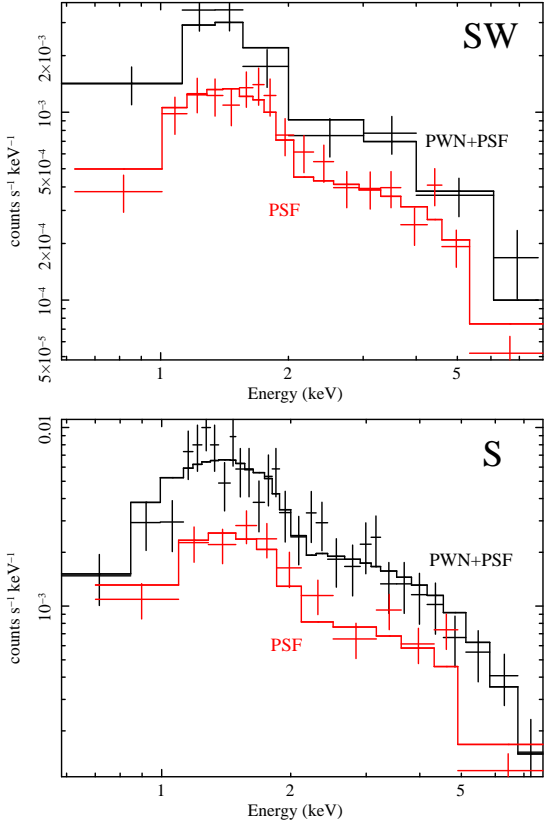

\centering
\includegraphics[scale=0.3,angle=-90]{f7a}
\includegraphics[scale=0.3,angle=-90]{f7b}
\caption{Spectra of the background-subtracted emission
in the SW ({\em top}) and S ({\em bottom}) regions (see Table 1 and Figure 1).
The red histograms show fits with the absorbed PL models of the
corresponding PSF wing spectra obtained in the ChaRT+MARX simulations.
The black histograms show fits of the observed (PSF+PWN) spectra with
the sum of two absorbed PL models,
one of which (with fixed parameters) describes the contribution of
the PSF wings.
}
\end{figure}

\subsubsection{Extended emission}
It is difficult to measure the spectrum of the alleged PWN
not only because of the small number of counts
but also because the PWN emission region overlaps with the wings of the 
pulsar PSF. 
We, however, attempted to estimate the spectral slope and the flux
of the 
PWN emission in the two regions shown in the bottom and top
panels of Figure~\ref{fig1}. First, using our ChaRT/MARX simulations, 
we extracted the PSF wing spectra in these regions and fit each of them 
with the absorbed PL model, freezing
the hydrogen column density at the best-fit value obtained for the pointlike
spectrum, $N_{\rm H, 21}=2.5$ (see Table~\ref{tab1} for the fitting parameters). 
These spectra are slightly harder than the 
PSF core spectrum because the PSF broadens with increasing energy. 
Extracting the background from the same regions that were used for the image analysis,
we then fit the spectra in the two regions by the sums of two absorbed PL 
models (PSF wings + PWN), with $N_{\rm H, 21}=2.5$ and
 the PSF wings model parameters frozen at
the previously obtained best-fit values. We see from Table~\ref{tab1} that the
spectra of the two regions have about the same
slopes, $\Gamma_{\rm pwn}\approx 1.6$,
statistically indistinguishable from the slope of the pointlike source,
 while the PWN flux in
the southern semi-annulus (S) is a factor of 3 higher than that in the
southwest rectangle (SW). The total flux and the luminosity of the 
alleged PWN are a factor of $\sim10$ lower than those of the pointlike source.

\section{Discussion}
\subsection{Pointlike source}
The pointlike source emission can include contributions from the pulsar,
the Be companion, and the unresolved part of the PWN created by the
colliding winds from the pulsar and the Be companion,
including the ultracompact intrabinary PWN, with
a size smaller than the separation between the binary components, 
$\lesssim 3 d_3^{-1}$ milliarcseconds at apastron.
Coronal emission from the Be companion, 
with an optically thin, thermal
plasma spectrum corresponding to a temperature of a few million kelvins,
might give some contribution
in low-energy channels, but, given the goodness of the 
single-component PL fit, the contribution of the coronal emission can be
neglected.
The pulsar emission could include a thermal component, emitted from
the neutron star surface, and a nonthermal component emitted from the
pulsar's magnetosphere. 
Since the observed spectrum excellently fits the single PL model, 
we conclude that the pulsar's thermal emission is too soft and faint to be seen
at the relatively large $N_{\rm H} \approx 2.5\times 10^{21}$ cm$^{-2}$. For instance, 
the thermal emission from the Geminga pulsar, whose spin-down
age is similar to that of B1259 and whose bulk surface
temperature is $\lesssim 5\times 10^5$ K  
\citep[see, e.g.,][]{Karga2005}, would be virtually undetectable at energies above 0.5 keV
if Geminga were at the distance of B1259.

For a typical X-ray efficiency
 of the magnetospheric emission, $\eta_{X,\rm psr}\equiv
L_{X,\rm psr}/\dot{E} \sim 10^{-4}$ \citep[see, e.g.,][]{Karga2008a}, one
would expect $L_{X,\rm psr}\sim 10^{32}$ erg s$^{-1}$, which is an order
of magnitude lower than the detected 
$L_{\rm 0.5-8\, keV} = 1.3\times 10^{33} d_3^2$ erg s$^{-1}$. 
However, as $\eta_{X,\rm psr}$
shows a large scatter \citep[from $<10^{-5}$ to $\sim 10^{-2}$ for pulsars with
similar values of $\dot{E}$ and $\tau$ --- see][]{Karga2009}, the
relatively high X-ray luminosity of B1259 does not rule out a substantial
contribution from the pulsar, especially taking into account that the
distance may be smaller than the assumed 3 kpc.
As the magnetospheric emission is expected to be strongly pulsed,
its contribution could be estimated by measuring X-ray pulsations
near apastron.
Such an attempt was made by \citet{Hirayama1999}
in observations with the {\sl ASCA} GIS detector, but the upper limit
on the pulsed fraction, $p_f \lesssim 40\%$,
was not  constraining.
To conclude, the pointlike emission likely includes contributions from the 
unresolved part of the PWN and
the pulsar's magnetosphere, with the total X-ray efficiency
$\eta_{\rm 0.5-8\, keV}= 1.6\times 10^{-3} d_3^2$.
Timing observation with {\sl XMM-Newton} or {\sl Chandra} could help
separate their contributions.

\subsection{Extended emission}
Thanks to the unprecedented angular resolution of {\sl Chandra}, our 
observation of B1259 has revealed, for the first time, extended X-ray emission
associated with this remarkable binary.
As the X-ray emission with a PL spectrum is usually generated by 
relativistic particles, and the pulsar is
the only natural source of such particles in B1259, we interpret the
extended emission as a PWN.

The detected extended 
PWN emission appears to consist of two components. Firmly detected
is only the relatively compact component south of the pulsar, 
with a characteristic size of 
$\sim 4''$ and luminosity of $\sim 10^{32}d_3^2$ erg s$^{-1}$.
The detection of the less luminous
 elongated component, of $\sim 15''$ length,
is less certain.

The elongated component could be interpreted as a pulsar jet.
Its $\sim 15''$ angular length 
corresponds to 
$\sim 7\times 10^{17}$ cm at $d=3$ kpc, which is comparable with the
$\sim 5\times 10^{17}$ cm and
$\sim 2\times 10^{17}$ cm projected lengths of the X-ray jets of
 the Vela and Geminga pulsars, respectively
\citep{Pavlov2003, Pavlov2010}. Its luminosity, $\sim 10^{31}$ erg s$^{-1}$,
is about $10^{-5}$ of the pulsar's spin-down power, versus $\sim 10^{-6}$ for
the Geminga and Vela pulsars.
To check this interpretation, longer {\sl Chandra} observations of B1259
are needed, at different binary phases. If the detected elongated feature
is indeed the pulsar jet, we do not expect substantial variations in its
luminosity and length, except may be for a short interval ($\sim 40$ d) around
periastron, when accretion of the Be companion wind might quench the pulsar
activity \citep{Wang2004}. 

The compact PWN component could be associated with
the shocked PW blown out of the binary
by the wind of the Be companion.
Similar to the case of a single pulsar moving
in the ISM  with a supersonic velocity, the collision of the two winds should create a bow shock within the binary (TA97) and 
a tail directed out of the binary approximately along the line 
connecting the Be companion with the pulsar\footnote{We assume here that
the pulsar's velocity (which varies from 12 km s$^{-1}$ at apastron to
176 km s$^{-1}$ at periastron) is much smaller than the
velocity of the Be companion's wind near the pulsar (see TA97).}.
As the pulsar moves along the binary orbit, the tail is 
expected to be bent in the 
direction opposite to the pulsar motion,
resembling an Archimedes spiral \citep{Dubus2006}. 
The tail's shape can described by the
equation 
\begin{equation}
R - R_p(\theta) =\frac{P_{\rm bin}(1-e^2)^{3/2}}{2\pi}
\int_\theta^{\theta_p} \frac{V_{\rm flow}(\theta')}{(1+e \cos\theta')^2} \, d\theta'\, , 
\end{equation}
where $R$ and $\theta$ are the radius and the azimuthal angle of a point
on the tail's central line in the reference frame centered at the Be
companion,
$R_p$ and $\theta_p$ are the radial coordinate and the true anomaly
 of the pulsar, 
and $V_{\rm flow}(\theta')$ is the flow speed that may vary along the tail.
The shape of the tail near apastron, 
when $|1+\cos\theta|\ll 1$ and $|1+\cos\theta_p|\ll 1$,
is given by the following
equation
\begin{eqnarray}
R - R_p(\theta)& \approx & \frac{P_{\rm bin} \bar{V}_{\rm flow}}{2\pi} \frac{(1+e)^{3/2}}{(1-e)^{1/2}}\, (\theta_p - \theta) \nonumber \\
& = & 2.3\times 10^{19} 
\frac{\bar{V}_{\rm flow}}{c} \frac{\theta_p-\theta}{2\pi}\, {\rm cm}\,,
\end{eqnarray}
where $\bar{V}_{\rm flow}$ is the average flow speed, and the numerical
estimate is given for the B1259 case 
($\theta_p = 1.0090 \pi$
 for our observation). 
The compact PWN in the ACIS image
 looks like a one-sided bulge on (or a ``blob'' attached to)
the pointlike source rather than a narrow tail. This is expected if
the angular size of the visible tail only slightly exceeds the size of the
PSF. In addition,
the tail may significantly
broaden by interactions with the ambient medium.

The luminosity of the compact component of the resolved PWN,
$L_{\rm 0.5-8\, keV}\sim 1.2\times 10^{32}d_3^2$ erg s$^{-1}$,
is a factor of $\sim 10$ lower than the luminosity of the pointlike source,
which includes the
PWN and pulsar contributions in an unknown proportion.
The corresponding X-ray efficiency, $\eta_{\rm 0.5-8\, keV}=
L_{\rm 0.5-8\, keV}/\dot{E}=1.4\times 10^{-4} d_3^2$,
is at the lower end of those observed in the tails of
supersonically moving solitary pulsars
\citep[see Table 5 in][]{Karga2008b}, but one should take into account
that a substantial part of the alleged tail of B1259 cannot be resolved even
with {\sl Chandra}, so the true tail luminosity and the X-ray efficiency may be
considerably larger than these estimates.

The observed angular size of the compact PWN, $\sim 4''$, 
which corresponds to the projected length
of $\sim 2\times 10^{17} d_3$ cm, 
and the fact that the PWN is seen only on one side of the pulsar,
put a constraint on the average flow speed: $\bar{V}_{\rm flow} \gtrsim 0.1c$.
Such high flow speeds 
have been obtained in numerical models for tails of bow-shock PWNe of
solitary pulsars 
(e.g., $V_{\rm flow}\sim 0.1c$--$0.3c$ in the narrow central channel of the
tail, and $\sim 0.8c$--$0.9c$ in the tail's ``sheath'', in the models by 
Bucciantini et al.\ 2005), but these speeds may become lower
at large distances from the pulsar because of the interaction
with the ambient medium (this effect was not included in the 
numerical models).
Observations of tails behind supersonically moving solitary pulsars
also show rather high flow speeds --- for instance, $\bar{V}_{\rm flow}
\gtrsim 0.02 c$ for
the 6 pc
tail behind PSR J1509--5850 \citep{Karga2008b}. 

Even higher flow speeds, corresponding to bulk motion Lorentz factors
$\gamma_{\rm bulk}$ up to $\sim 100$, were obtained by \citet{Bogovalov2008} in hydrodynamical
simulations of
flows produced by the collision of the relativistic (pulsar) and
nonrelativistic (Be star) winds in a binary system. The ultrarelativistic
flows are formed when the ratio $\eta$ of the momentum flux from the
pulsar to that from the Be star exceeds $1.25\times 10^{-2}$, in which
case the TS surface becomes ``unclosed'' (i.e., the  
back surface of the
TS bullet disappears). 
In the case of B1259,
 $\eta$ may vary between $\sim 10^{-2}$ and $\sim 1$ along the binary orbit,
being as large as 0.1--1 around apastron (when the PW interacts
with the so-called polar component of the Be star outflow). If the 
confined flow indeed reaches such high 
speeds, it can penetrate farther into the ambient
medium, which makes the suggested interpretation of the observed
 asymmetric emission even more plausible. 
\citet{Dubus2010} note
that 
the properties of the observed X-ray and VHE emission from B1259
are not compatible with 
the strong Doppler boosting corresponding to such large $\gamma_{\rm bulk}$.
However, it is possible that, while the bulk of the observed high-energy
emission is generated within (or very close to) the binary, 
where the flow is only mildly
relativistic, the extended emission is generated by electrons supplied
by the accelerated flow that slows down at large distances from the binary.

The interpretation of the compact PWN as a shocked PW outside the binary
can constrain the wind paramenters in various models of B1259.
The X-ray PWN emission could be due to synchrotron radiation of 
ultrarelativistic electrons, accelerated at the inner shock front within
the binary (TA97).
To generate X-ray photons, $E\gtrsim 0.5$ keV, the electron energy
$\gamma mc^2$ and the magnetic field $B$ should be high enough:
$B \gamma^2 \gtrsim 10^{11}$ 
(where $B$ is in units
of gauss). On the other hand, the cooling time, $t_{\rm cool}$, for
the relativistic electrons in the flow should be longer than
the flow time, $t_{\rm flow} \sim R/\bar{V}_{\rm flow}
\sim 3\times 10^7 R_{17} V_{9.5}^{-1}$ s, where 
$R_{17} = R/(10^{17}\, {\rm cm})$, $V_{9.5} = \bar{V}_{\rm flow}/(10^{9.5}\, 
{\rm cm}\, {\rm s}^{-1})$.
Generally, both the synchrotron cooling in the magnetic field
 and IC 
cooling
in the radiation field of the Be companion can be important.
At the very high energies of electrons emitting synchrotron X-rays,
$\gamma \gg mc^2/\epsilon \sim 10^5 $
(where $\epsilon \sim 3 kT_{\rm eff} \sim 5$ eV is the mean photon energy),
the IC cooling occurs in the Klein-Nishina regime.
The IC cooling rate in this regime is much lower 
than in the Thomson regime: 
$\dot{\gamma}_{\rm IC, KN}/\dot{\gamma}_{\rm IC,T}
\sim (mc^2/\gamma\epsilon)^2\ln(\gamma\epsilon/mc^2)\sim 10^{-4}\,\gamma_7^{-2}
(\ln \gamma_7 +5)$, and it is much lower than
the synchrotron cooling rate beyond the binary orbit. 
Therefore, the cooling time is determined by synchrotron radiation,
$t_{\rm cool}\sim t_{\rm syn} \sim 5\times 10^8 \gamma^{-1} B^{-2}$ s, and it exceeds
the flow time 
at $\gamma B^2 < 16 V_{9.5} R_{17}^{-1}$.
Taking into account the condition $B\gamma^2 \gtrsim 10^{11}$ derived above,
we obtain
\begin{eqnarray}
B & < & 1.4\times 10^{-3}\, V_{9.5}^{2/3} R_{17}^{-2/3}\,\, {\rm G}, \nonumber\\
\gamma & > & 0.9 \times 10^7\, V_{9.5}^{-1/3} R_{17}^{1/3}\, ,
\end{eqnarray}
for the typical magnetic field and Lorentz factor.
The upper limit on magnetic field is much lower than the magnetic field 
in the ultracompact intrabinary PWN (e.g., $B\sim 1$ G 
in the TA97 model), while the lower limit on Lorentz factor is close
to $\gamma_{\rm max}$ of the electron energy spectrum in the shocked
(uncooled) PW within the binary (TA97). 
To reconcile such high energies and low magnetic fields with the 
intrabinary PWN models, one has to
assume that the radiating
electrons are additionally accelerated in the tail, in some internal shocks
or via magnetic field reconnection that transforms the magnetic field
energy into the kinetic energy of particles.

As noted by Chernyakova et al.\ (2006, 2009), the TA97 model cannot explain
some of the observed properties of the B1259's radiation (such as the 
hardening of the X-ray spectrum near periastron). These authors \citep[see also][]{Neronov2007} discuss another
model \citep[first considered by][]{Chernyakova1999},
 in which the X-ray emission is
generated by the IC scattering of stellar photons off moderately relativistic
electrons with Lorentz factors $\gamma\sim 10$--100 (much lower than in the TA97 model).
As such scattering occurs in the Thomson
regime, the cooling rate is 
$\dot{\gamma}_{\rm IC}=-(4/3)(\sigma_{\rm T}c/mc^2)\gamma^2 U_{\rm ph}$,
where
$U_{\rm ph}=L_*/(4\pi c R^2) = 1.7\times 10^{4} L_{*,38} R_{17}^{-2}$ eV cm$^{-3}$
is the photon energy density at the distance $R =10^{17} R_{17}$ cm from the Be companion 
(approximately equal to the distance from the binary at $R\gg a$),
and $L_* = 10^{38} L_{*,38}$ erg s$^{-1}$ is the luminosity of the stellar
companion. 
At realistic values of the magnetic field, 
$B\lesssim 800 L_{*,38} R_{17}^{-1}$ $\mu$G,
the photon energy density exceeds the magnetic energy density,
so the IC cooling is more efficient. 
Taking into account that $\dot{\gamma}_{\rm IC} \propto R^{-2}$,
the electron's energy at distance $R$
is given by the equation
\begin{equation}
\gamma (R) = \gamma_0\left[1 +\frac{\gamma_0}{\gamma_c}\left(1-\frac{R_0}{R}\right)\right]^{-1}\, ,
\end{equation}
where $\gamma_0=\gamma(R_0)$,
 $\gamma_c = 3\pi m c^2 R_0 V_{\rm flow}/(\sigma_{\rm T}L_*)\approx 
3.7\times 10^4 R_{0,14} V_{9.5} L_{*,38}^{-1}$. 
If we assume $\gamma_0 \sim 10$--100 close to the binary orbit at apastron 
($R_0\sim
10^{14}$ cm), then $\gamma_0/\gamma_c \sim (3$--$30)\times 10^{-4} V_{9.5}^{-1}$,
which means that
the total relative energy loss by an electron, 
$(\gamma_0 - \gamma_\infty)/\gamma_0 =
(\gamma_0/\gamma_c)[1+(\gamma_0/\gamma_c)]^{-1} \approx \gamma_0/\gamma_c$, 
is very small
at realistic $V_{\rm flow}\gg 10$--100 km s$^{-1}$. 
Moreover, the upper limit on the X-ray efficiency of the resolved PWN
between distances $R_1$ and $R_2$ from the binary,
$[\gamma(R_1) - \gamma(R_2)]/\gamma_0 \sim
(\gamma_0/\gamma_c)R_0 (R_1^{-1} - R_2^{-1})
\sim 10^{-6} (\gamma_0/30) V_{9.5}^{-1}\left(R_{1,17}^{-1}-R_{2,17}^{-1}\right)$, 
is much smaller than
the observed value, $\sim 10^{-4}$. 
This means that the observed extended X-ray emission cannot be explained
by the IC scattering, and the synchrotron emission of electrons with
TeV energies is a more plausible interpretation. 

The suggested interpretation of the compact PWN as a bent tail of the
shocked PW blown out of the binary by the wind of the Be companion
implies that the compact PWN rotates around
the pointlike source with the binary period, and its size and
luminosity should depend on binary phase. This can
be verified by observing this system with {\sl Chandra} at different
binary phases. 

If these observations show the compact PWN  at the same position
with respect to the pulsar, another interpretation would be required
to explain the extended emission. In principle, it might be a tail or trail
of shocked PW behind the moving binary blown out by
 the head-on wind of the ambient ISM 
matter. The proper motion of B1259 has not been measured yet, but 
the speed of the high-mass binary should be much lower than that of
a solitary pulsar, so that it would be hard to explain even so short tail/trail 
as caused by the motion of B1259 in the ISM. 

Finally, we note that strong evidence of 
extended emission around
another $\gamma$-ray binary, the microquasar LS 5039, has been recently
found in a {\sl Chandra} ACIS observation \citep{Durant2011}.
It shows that the compact object in LS 5039 is likely a rotation-powered
pulsar rather than an accreting neutron star or a black hole.
It also means that B1259 is not the only high-mass X-ray binary surrounded
by an extended PWN.

\acknowledgments
We thank Nicholas Lee and the CXC Calibration team for the discussion of the 
mirror misalignment effect
on ACIS images, and Maria Chernyakova for the discussion of the {\sl XMM-Newton}
spectrum of B1259 near apastron.
We also thank the anonymous referee, whose remarks has allowed us to improve
the presentation.
Support for this work was provided by the National Aeronautics and Space
Administration through {\sl Chandra} Award Number GO9-0070X issued
by the {\sl Chandra} X-ray Observatory Center, which is operated by the
Smithsonian Astrophysical Observatory for and on behalf of the National
Aeronautics Space Administration under contract NAS8-03060. The work 
was also partially supported by NASA grants NNX09AC84G and NNX09AC81G,
and NSF grants No.\ 0908733 and 0908611. The work by GGP was partly
supported by the Ministry of Education and Science of the Russian
Federation (Contract No.\ 11.G34.31.0001).


\begin{table}
\caption{Properties of X-ray Emission from B1259
\label{tab1}}
\begin{tabular}{lcccccc}\hline\hline
Region\tablenotemark{a}&$N_{\rm H,21}$&$\Gamma$&${\mathcal N}_{-6}$\tablenotemark{b}& 
$F_{-14}$\tablenotemark{c}&$F^{\rm unabs}_{-14}$\tablenotemark{d}  &$L_{32}$\tablenotemark{e} \\
\hline
Pointlike&2.50$^{+0.60}_{-0.54}$&1.64$^{+0.12}_{-0.11}$&171$^{+26}_{-22}$& 
$83.0\pm4.5$ &119$^{+6}_{-7}$ & 13 \\
\hline
PSF S&\multirow{2}{*}{[2.5]}&1.37$^{+0.18}_{-0.17}$&10$\pm$2 &$\ldots$&$\ldots$& $\ldots$ \\
PWN S& &1.57$^{+0.29}_{-0.27}$&17$\pm$4& 
$9\pm 1$ &11$\pm$2 & 1.2 \\
\hline
PSF SW&\multirow{2}{*}{[2.5]}&$1.34^{+0.15}_{-0.15}$&5.1$^{+0.8}_{-0.7}$ &$\ldots$&$\ldots$ & $\ldots$ \\
PWN SW& &1.65$^{+0.56}_{-0.49}$&7.3$^{+3.4}_{-2.7}$& 3$\pm$1&4$\pm$1 & 0.4 \\
\hline\hline
\end{tabular}
\tablecomments{The errors shown represent 90\% confidence intervals. The fluxes and luminosities are for the 0.5--8 keV band.
} \\
\tablenotetext{a}{The region marked S is the semiannulus $0\farcs75 < r < 4''$
in the semicircle south of the pointlike source, 
shown in the bottom page of Figure 1. The region marked SW is the southwest
part, $1''<l<18''$, of the $36''\times 5''$ rectangle shown in the upper
panel of Figure 1.}
\tablenotetext{b}{ PL normalization in units of 10$^{-6}$ photons keV$^{-1}$ cm$^{-2}$ s$^{-1}$ at 1 keV.} \\
\tablenotetext{c}{Observed (absorbed) flux in units of 10$^{-14}$ erg cm$^{-2}$ s$^{-1}$ (in $1''$ radius aperture for the pointlike source, which contains 86\% energy fraction).} \\
\tablenotetext{d}{Unabsorbed flux in units of 10$^{-14}$ erg cm$^{-2}$ s$^{-1}$(aperture-corrected for the pointlike source).}\\
\tablenotetext{e}{~Luminosity in units of $10^{32} d_3^2$ erg s$^{-1}$.}
\end{table}

\end{document}